\documentclass[12pt,a4paper]{article}
\usepackage{latexsym}
\usepackage{epsf}
\makeatletter
\@addtoreset{equation}{section}
\makeatother

\begin{document}

\begin{flushright}
CERN--TH/97--125\\
UG-5/97\\
{\bf hep-th/9706117}\\
June  $15$th, $1997$
\end{flushright}

\begin{center}

%title

{\large {\bf Kaluza-Klein Monopoles and Gauged Sigma-Models}}

\vspace{.9cm}

%authors
{\large
{\bf Eric Bergshoeff}
\footnote{E-mail address: {\tt E.Bergshoeff@phys.rug.nl}}
{\bf and Bert Janssen}
\footnote{E-mail address: {\tt B.Janssen@phys.rug.nl}}\\
{\it Institute for Theoretical Physics\\
University of Groningen\\
Nijenborgh 4, 9747 AG Groningen\\
The Netherlands}\\
}
\bigskip

{\large
{\bf Tom\'as Ort\'{\i}n}
\footnote{E-mail address: {\tt Tomas.Ortin@cern.ch}}${}^{,}$
\footnote{Address after October 1997:  {\it IMAFF, CSIC, 
Calle de Serrano 121, E-28006-Madrid, Spain}}\\
\vspace{.4cm}
{\it C.E.R.N.~Theory Division}\\
{\it CH--1211, Gen\`eve 23, Switzerland}\\
}

\vspace{.8cm}

%%%%%%%%%%%%%%%%%%%%%%%%%%%%%%%%%%%%%%%%%%%%%%%%%%%%%%%%%%%%%%%%%%%%%%

{\bf Abstract}

\end{center}

\begin{quotation}

\small

We propose an effective action for the eleven-di\-men\-sio\-nal
(bosonic) Kaluza-Klein monopole solution. The construction of the
action requires that the background fields admit an Abelian isometry
group.  The corresponding sigma-model is gauged with respect to this
isometry.  The gauged sigma-model is the source for the monopole
solution.  A direct (double) dimensional reduction of the action leads
to the effective action of a 10-dimensional D-6-brane (IIA
Kaluza-Klein monopole). We also show that the effective action of the
10-dimensional heterotic Kaluza-Klein monopole (which is a truncation
of the IIA monopole action) is T-dual to the effective action of the
solitonic 5-brane.  We briefly discuss the kappa-symmetric extension
of our proposal and the possible role of gauged sigma-models in
connection with the conjectured M-theory 9-brane.

\end{quotation}

\begin{flushleft}
CERN--TH/97--125\\
\end{flushleft}

\newpage

\pagestyle{plain}

%%%%%%%%%%%%%%%%%%%%%%%%%%%%%%%%%%%%%%%%%%%%%%%%%%%%%%%%%%%%%%%%%%%%%%

\section*{Introduction}

Eleven-dimensional supergravity \cite{eleven} is believed to describe
the low-energy behaviour of (uncompactified) M-theory \cite{townsend},
which may be a theory of supermembranes \cite{be1}.  To gain a better
understanding of M-theory it is therefore of interest to study the
different kinds of solutions to the supergravity equations of motion.
Of particular interest are those solutions that, upon reduction to ten
dimensions, lead to the D-p-brane solutions of IIA supergravity
\cite{kn:DKL}. By now, the 11-dimensional origin of the D-p-brane
solutions (with $p$ even) are well understood with the exception of
the D-8-brane whose 11-dimensional interpretation is still a
mystery (see, however, the discussion).

It turns out that 11-dimensional supergravity admits two ``brane''
solutions, the M-2-brane and M-5-brane. Their reduction leads to the
D-2-brane and D-4-brane solution, respectively.  There is no such
brane interpretation for the D-0-brane and D-6-brane. These solutions
are related to a purely gravitational Brinkmann wave (W11) and a
Kaluza-Klein monopole (KK11) in eleven dimensions \cite{to1}.  The
metric corresponding to these solutions does not split up into
isotropic worldvolume and transverse directions and therefore does not
describe a standard brane. Nevertheless, due to their 10-dimensional
D-brane interpretation, these solutions are expected to play an
important role in M-theory. It is therefore of interest to get a
better understanding of these solutions and the role they play in
M-theory.

A natural question to ask is, what is the effective action
corresponding to the Brinkmann wave and Kaluza-Klein monopole? Such
effective actions occur as source terms in the supergravity equations
of motion.  The effective action of the Brinkmann wave is a massless
particle (or any other kind of massless $p$-brane \cite{prep}) moving
at the speed of light. The zero modes of the 11-dimensional
Kaluza-Klein monopole have been recently discussed in \cite{kn:H}. In
this letter we will make a concrete proposal for the KK11 effective
action that involves those zero modes. To motivate our proposal we
first show that the purely gravitational part\footnote{A precise
  definition will be given in the next section.} of the action is the
source of the KK11 solution.  Next, we show that, up to the yet
undetermined Wess-Zumino term, a direct dimensional reduction of the
KK11 action leads to the effective action of a 10-dimensional
D-6-brane, while a double dimensional reduction leads to the effective
action for a IIA Kaluza-Klein monopole (KK10A). In the latter
reduction we only consider the purely gravitational part of the
action.  Finally, we will show that the effective action of the
10-dimensional heterotic Kaluza-Klein monopole (KKh), which is a
truncation of the KK10A action, is T-dual to the effective action of
the heterotic solitonic 5-brane (P5h), again up to the WZ term
\cite{prep}.

%%%%%%%%%%%%%%%%%%%%%%%%%%%%%%%%%%%%%%%%%%%%%%%%%%%%%%%%%%%%%%%%%%%%%%

\section{The KK11 Effective Action}
\label{sec-pointkk}

Our starting point is the 11-dimensional KK11 solution 
\cite{kn:SGP}\footnote{The discussion below can easily be extended to
arbitrary dimensions $d$. For simplicity we restrict ourselves to
d=11.}

\begin{equation}
ds^{2}= \eta_{ij}dy^{i}dy^{j} 
-H^{-1}\left(dz +{A}_{m} dx^{m}\right)^{2} 
-H (dx^{m})^2\, ,
\label{eq:KK}
\end{equation}

\noindent where $i,j=0,\ldots,6$, $m,n=7,8,9,$ and
 $z=x^{10}$ and where

\begin{equation}
{F}_{mn}= 2\partial_{[m}{A}_{n]} 
= \epsilon_{mnp}\partial_{p} H\, ,
\hspace{1cm}
\partial_{m}\partial_{m}H=0\, .
\end{equation}

The solution (\ref{eq:KK}) has $8$ isometries and therefore it
represents an extended object. At first sight one might think that the
solution represents a 7-brane (with non-isotropic worldvolume
directions) but it turns out that the isometry in the direction $z$ is
special and cannot be interpreted as a worldvolume direction
\cite{kn:H}. We are therefore dealing with a 6-brane, with a
7-dimensional worldvolume, that has an additional isometry in one of
the 4 transverse directions.

The KK11 solution preserves half of the supersymmetry and must
correspond, after gauge fixing, to a 7-dimensional supersymmetric
field theory.  The natural candidate for such a field theory involves
a vector multiplet with 3 scalars and one vector \cite{kn:H}. We are
now faced with a dilemma. Since the KK11-monopole moves in 11
dimensions, we have 11 embedding coordinates.  Fixing the
diffeomorphisms of the 7-dimensional worldvolume we are left with 4
instead of 3 scalars. At this point one might argue that, to eliminate
the extra scalar d.o.f., we need an extra diffeomorphism, i.e.~an
8-dimensional worldvolume, but this would upset the counting of the
worldvolume vector components.  We therefore need a new mechanism to
eliminate the unphysical scalar degree of freedom.  As we will see
below, this can be done by gauging an Abelian isometry in the
effective sigma-model, i.e.~we propose to work with a {\sl gauged
  sigma-model}.

A second characteristic feature of our proposal is that the
``KK11-brane'' couples to a scalar $k$ constructed from the Killing
vector $k^{\mu}$ that generates the isometry we are gauging. The
coupling manifests itself as a factor $k^{2}$ in front of the kinetic
term in the effective action.  Since, in coordinates adapted to the
isometry, $g_{zz}=-k^{2}$ and the length of the $z$-dimension is

\begin{equation}
2\pi R_{z}= \int dz |g_{zz}|^{1/2}= \int dz k\, , 
\end{equation}

\noindent the tension of the KK11-brane (and of the KK brane in any dimension) 
is proportional to $R_{z}^{2}$.

To be concrete, we propose the following expression for the kinetic
term of the KK11 effective action (a proper WZ term needs to be added 
\cite{prep}):

\begin{equation}
S_{{\rm KK11}} = - T_{{\rm KK11}} 
\int d^7\xi\  k^2 \sqrt {|{\rm det}\
(\partial_i X^\mu\partial_j X^\nu \Pi_{\mu\nu} + k^{-1}{\cal F}_{ij})|}\, ,
\end{equation}

\noindent where  $k^\mu$ is the Killing vector
associated to the isometry direction $z$ and 

\begin{equation}
k^2 = -k^\mu k^\nu g_{\mu\nu}\, .
\end{equation}

Furthermore,

\begin{equation}
\left\{
\begin{array}{rcl}
\Pi_{\mu\nu} & = & g_{\mu\nu} + k^{-2} k_\mu k_\nu\, ,\\
& & \\
{\cal F}_{ij} &=& \partial_i V_j - \partial_j V_i - k^\mu
\partial_i X^\nu\partial_j X^\rho C^{(3)}_{\mu\nu\rho}\, . \\
\end{array}
\right.
\end{equation}

\noindent The field $C^{(3)}$ is the 3-form potential of 11-dimensional
supergravity.  

Observe that the components of the ``metric'' $\Pi_{\mu\nu}$ in the
directions of $k^{\mu}$ vanish: 

\begin{equation}
k^{\mu} \Pi_{\mu\nu}=0\, . 
\end{equation}

\noindent $\Pi_{\mu\nu}$ is effectively a 10-dimensional metric and, in
coordinates adapted to the isometry generated by $k^{\mu}$, the
coordinate $z$ and the corresponding field $Z(\xi)$ associated to this
isometry simply do not occur in the action.

Sometimes, we will consider in our discussion only the purely
gravitational part of the KK11 action, i.e.~we will set the
worldvolume vector field strength ${\cal F}_{ij}$ equal to zero and
we will ignore the WZ term:

\begin{equation}
\label{NG}
S^{\rm grav.}_{{\rm KK11}} = - T_{{\rm KK11}} 
\int d^7\xi\  k^2 \sqrt {|{\rm det}\
(\partial_i X^\mu\partial_j X^\nu \Pi_{\mu\nu}) |}\, .
\end{equation}

This part of the action can be written using an auxiliary
worldvolume metric $\gamma_{ij}$ in the Howe-Tucker form

\begin{equation}
\label{P}
S^{\rm grav.}_{{\rm KK11}} = - {T_{{\rm KK11}}\over 2} 
\int d^{7}\xi  
\sqrt{|\gamma|} \left[k^{4/7} \gamma^{ij}
\partial_{i}X^{\mu} \partial_{j}X^{\nu}\Pi_{\mu\nu}-5\right]\, .
\end{equation}
Eliminating $\gamma_{ij}$ from (\ref{P}) leads to the expression
given in (\ref{NG}). An alternative form of (\ref{P}),
which makes the relation with a gauged sigma-model clear,
is obtained using an auxiliary
worldvolume vector field $C_i$:

\begin{equation}
\label{GS}
S^{\rm grav.}_{{\rm KK11}} = - {T_{{\rm KK11}}\over 2} 
\int d^{7}\xi  
\sqrt{|\gamma|} \left[k^{4/7} \gamma^{ij}
D_{i}X^{\mu} D_{j}X^{\nu}g_{\mu\nu}-5\right]\, ,
\end{equation} 
with the covariant derivative defined as 

\begin{equation}
D_i X^\mu = \partial_i X^\mu + C_i k^\mu\, .
\end{equation}

\begin{figure}[!ht]
\begin{center}
\leavevmode
\epsfxsize= 13cm
\epsffile{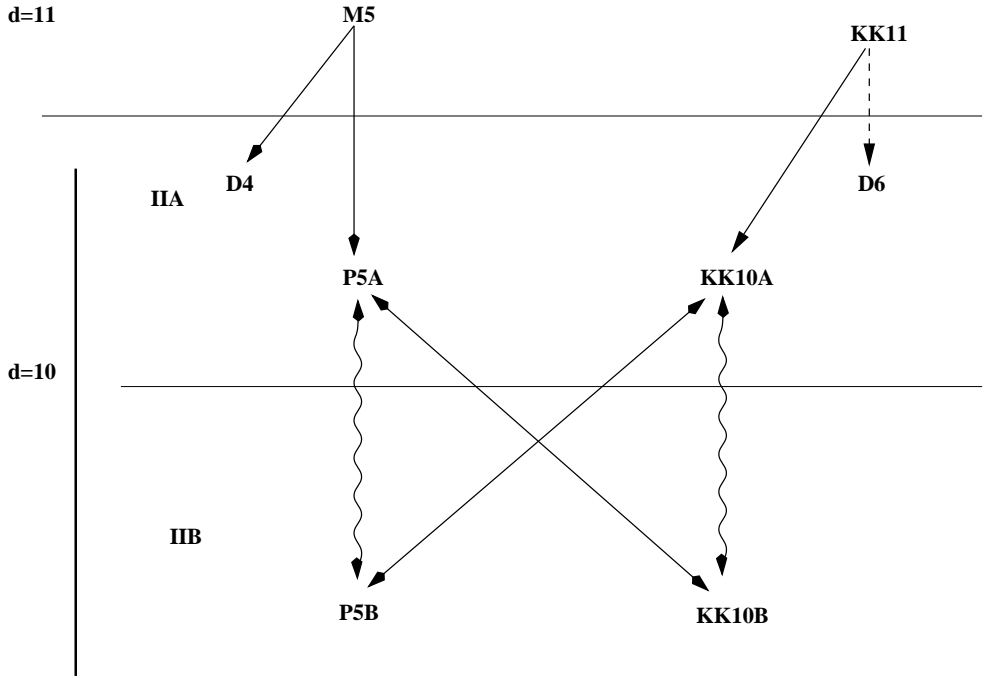}
\caption{{\scriptsize
Relation between solutions and effective actions of KK
  monopoles and other extended objects in M~theory and type~II string
  theories. Vertical arrows indicate direct dimensional reduction.
  Vertical dashed lines indicate direct dimensional reduction in the
  special direction $z$.  Oblique arrows indicate double dimensional
  reduction.  Double arrows indicate T~duality relations and wiggly
  lines indicate that the supergravity solutions are identical,
  although the effective actions are not.}}
\label{fig:KKdual}
\end{center}
\end{figure}

In the remaining part of this letter we will collect evidence in
favour of our proposal. We first show that the KK11 action is the
source for the KK11 solution. Next, we compare the KK11 action with
the kinetic terms of other known actions via dimensional reduction and
T-duality. The relations between the KK11 action of M-theory and the
different actions of the IIA/IIB theories are given in
Figure~\ref{fig:KKdual}.  The worldvolume fields of the corresponding
actions are given in Table~\ref{tab-degrees}.

\begin{table}
\footnotesize
\begin{center}
\begin{tabular}{||c||c|c|c||c||}
\hline\hline
& & & & \\
Object & Worldvolume & Fields                   & $\#$\ of d.o.f.  &Total\\
& & & & \\
\hline\hline
      &     & $X^\mu$        & $11-7$ ``$-1$''$= 3$    &      \\
KK11  & 6+1 &                &                         &  8   \\
      &     & $V_i$          & $7-2        = 5$        &      \\
\hline\hline
      &     & $X^\mu$        & $10-7       = 3$        &      \\
D6    & 6+1 &                &                         &  8   \\
      &     & $V_{i}$        & $7-2        = 5$        &      \\
\hline\hline
      &     & $X^\mu$        & $10-6$ ``$-1$''$= 3 $   &      \\
KK10A & 5+1 & $V_{i}$        & $6-2        = 4$        &  8   \\
      &     &  $S$           &     1                   &      \\
\hline\hline
      &     & $X^\mu$        & $10-6       = 4$        &      \\
P5B   & 5+1 &                &                         &  8   \\
      &     & $W_{ijk}$      & $4$                     &      \\
\hline\hline
      &     & $X^\mu$        & $11-6 = 5$              &      \\
M5    & 5+1 &                &                         &  8   \\
      &     & $V^{+}_{ij}$   & $3$                     &      \\
\hline\hline
      &     & $X^\mu$        & $10-6 = 4$              &      \\
P5A   & 5+1 & $V^{+}_{ij}$   & $3$                     & 8    \\
      &     &  $S$           &     1                   &      \\
\hline\hline
      &     & $X^\mu$        & $10-6$ ``$-1$''$= 3$    &      \\
      &     & $V^{+}_{ij}$   & $3$                     &      \\
KK10B & 5+1 &                &                         &  8   \\
      &     &  $S$           &     1                   &      \\
      &     &  $T$           &     1                   &      \\
\hline\hline
\end{tabular}
\end{center}
\caption[Table of worldvolume fields and scalar degrees of freedom]
{{\scriptsize The table gives the worldvolume fields and number of degrees of
freedom of the different objects that appear in Figure~\ref{fig:KKdual}.
The ``-1'' in the fourth column indicates that a scalar degree of freedom
is eliminated by gauging an Abelian isometry.}}
\label{tab-degrees}
\end{table}

%%%%%%%%%%%%%%%%%%%%%%%%%%%%%%%%%%%%%%%%%%%%%%%%%%%%%%%%%%%%%%%%%%%%

\section{The KK11 Action as Source Term}

We first demonstrate that the KK11 action is the source of the
11-dimensional KK monopole\footnote{Results concerning singular
sources in General Relativity are known to be highly
coordinate-dependent. In particular, in isotropic coordinates the
sources of extremal black holes and branes seem to be placed at the
horizon, which is non-singular but looks like a point in this
coordinate system. The reason for this is that the coordinates chosen
do not cover the region where the physical singularity is and the flux
lines of the different fields seem to come out of the point that
represents the horizon. Our results have to be understood in the same
sense. We thank G.W.~Gibbons and P.K.~Townsend for discussions on this
point.}. For simplicity, we only consider the purely gravitational
part. This leads to the following bulk plus source term:

\begin{equation}
\begin{array}{rcl}
S & = & {1\over \kappa} \int d^{11}x\sqrt{|g|}[R] \\
& & \\
& & 
-{T_{KK11}\over 2}\int d^{7}\xi  
\sqrt{|\gamma|} \left[k^{4/7} \gamma^{ij}
\partial_{i}X^{\mu} \partial_{j}X^{\nu}\Pi_{\mu\nu}-5\right]\, , \\
\end{array}
\end{equation}

\noindent where $\kappa=16\pi G^{(11)}_{N}$. 

To determine the Einstein equations
we have to take into account the metric factors hidden in factors like
$k^{2}$. The rule is that $k^{\mu}$ is independent of the metric.
The metrics that have to be varied in $k^{2}$ and $\Pi$ are
shown explicitly below:

\begin{equation}
\left\{
\begin{array}{rcl}
k^{2} & = & -k^{\mu}k^{\nu}g_{\mu\nu}\, ,\\
& & \\
\Pi_{\mu\nu} & = & g_{\mu\nu} 
+\left(k^{\alpha}k^{\beta}g_{\alpha\beta} \right)^{-1} k^{\rho}g_{\rho\mu}
k^{\sigma}g_{\sigma\beta}\, .\\
\end{array}
\right.
\end{equation}

The equations of motion for $g_{\mu\nu}$ and $X^{\nu}$ are

\begin{equation}
\begin{array}{rcl}
G^{\alpha\beta} +{T_{KK11}\kappa \over 2\sqrt{|g|}} \int d^{7}\xi 
\sqrt{|\gamma|} k^{4/7}\gamma^{ij} \left\{ -\frac{2}{3}
k^{-2}k^{\alpha}k^{\beta} \Pi_{ij} 
+\partial_{i}X^{\alpha}\partial_{j}X^{\beta}\right. & &  \\
& & \\
\left.
-2k^{-3}k^{\alpha}k^{\beta}k_{i}k_{j} 
-2k^{-2} k^{(\alpha}\partial_{i}X^{\beta)}k_{j}\right\}
\delta^{(10)}(x-X) & = & 0\, , \\
& & \\
\tilde{\nabla}^{2}X^{\rho} +\tilde{\Gamma}_{\mu\nu}{}^{\rho}
\partial_{k}X^{\mu}\partial^{k}X^{\nu} & = & 0\, ,\\
\end{array}
\end{equation}

\noindent where $\tilde{\Gamma}$ are the Christoffel symbols of the 
``metric''

\begin{equation}
\tilde{g}_{\mu\nu}= k^{2}\Pi_{\mu\nu}\, ,
\end{equation}

\noindent and $\tilde{\nabla}^{2}$ is the Laplacian with respect
to the worldvolume metric

\begin{equation}
\tilde{\gamma}_{ij}=k^{-4/7}\gamma_{ij}\, ,
\end{equation}

\noindent while the equation of motion for $\gamma_{ij}$ simply
implies

\begin{equation}
\gamma_{ij}=k^{4/7}\Pi_{ij}\, ,
\hspace{1cm}
\tilde{\gamma}_{ij}=\Pi_{ij}\, .
\end{equation}

Observe that in the source term for the Einstein equation, instead of
the usual $11$-dimensional Dirac delta function $\delta^{(11)}$ we
have written a $10$-dimensional Dirac delta function $\delta^{(10)}$.
The reason is that the KK monopole effective action does not depend on
all the coordinates, as we have seen. In adapted coordinates, it will
not depend on $Z$, and, therefore, a factor of $\delta (z-Z)$ is
absent.

We find that the only non-vanishing components
of the Einstein tensor with upper indices for the KK monopole metric
(\ref{eq:KK}) are

\begin{equation}
\left\{
\begin{array}{rcl}
G^{zz} & = & -H^{-1}\partial^{2}H\, ,\\
& & \\
G^{ij} & = & \frac{1}{2}\eta^{ij}H^{-2}\partial^{2}H\, .
\end{array}
\right.
\label{eq:Einstensor}
\end{equation}

\noindent For the embedding coordinates we make the following ansatz:

\begin{equation}
X^{i}=\xi^{i}\, ,
\hspace{1cm}
Z=X^{m}  =0\, ,
\end{equation}

\noindent which justifies that the indices $i,j$ can be used both as
worldvolume indices as well as the first 7 target-space indices.  The
above ansatz tells us that the extended object worldvolume lies in the
position $x^{m}=z=0$.  With this ansatz
$\tilde{\gamma}_{ij}=\eta_{ij}$ and the equation of motion for the
scalars $X^{\mu}$ is satisfied.

Taking into account the form of the metric given in Eq.~(\ref{eq:KK})
we find

\begin{equation}
k^{\mu}=\delta^{\mu z}\, ,
\hspace{.5cm}
k_{i}=0\, ,
\hspace{.5cm}
\Pi_{ij}=g_{ij}=\eta_{ij}\, .
\end{equation}

\noindent At this stage we find that most of the components of 
the Einstein equation are automatically satisfied. The only
non-trivial ones are the $zz-$ and $ij-$components. Using
Eqs.~(\ref{eq:Einstensor}) we find that both lead to the same
equation:

\begin{equation}
\partial^{2}H= -T_{KK11}\ \kappa \delta^{(3)}(\vec{x})\, , 
\end{equation}

\noindent which is solved by

\begin{equation}
H=1+{T_{KK11} \kappa\over 4\pi}\frac{1}{|\vec{x}|}\, .
\end{equation}

We thus conclude that the KK11 action is indeed the source of 
the 11-dimensional Kaluza-Klein monopole.

%%%%%%%%%%%%%%%%%%%%%%%%%%%%%%%%%%%%%%%%%%%%%%%%%%%%%%%%%%%%%%%%%%%%

\section{Dimensional Reduction}

In this section we will first show that the direct dimensional
reduction of the KK11 action leads to the D-6-brane action and next
that the double dimensional reduction of the (purely gravitational
part of the) same action leads to the KK10A effective action. In this
section we will indicate 11-dimensional (10-dimensional) fields
by double (single) hats.

%%%%%%%%%%%%%%%%%%%%%%%%%%%%%%%%%%%%%%%%%%%%%%%%%%%%%%%%%%%%%%%%%%%%

\subsection{Direct Dimensional Reduction}

To perform the direct dimensional reduction it is convenient to use
coordinates adapted to the gauged isometry. In such a coordinate
system we have

\begin{equation}
{\hat {\hat {k}}}^{\hat {\hat {\mu}}} = \delta^{{\hat {\hat \mu}}z}\, ,
\end{equation}

\noindent and the only non-vanishing components of
$\hat{\hat{\Pi}}_{\hat{\hat{\mu}}\hat{\hat{\nu}}}$ are

\begin{equation}
\hat{\hat{\Pi}}_{\hat{\mu}\hat{\nu}} 
=  \hat{\hat{k}}^{-1} \hat{g}_{\hat{\mu}\hat{\nu}}\, .
\end{equation}

\noindent Furthermore

\begin{equation}
\hat{\hat{g}}_{zz} = - \hat{\hat{k}}^{2}= -e^{\frac{4}{3}\hat{\phi}}\, .  
\end{equation}
Substituting the latter two equations into the KK11 action gives 

\begin{equation}
S_{{\rm D6}} = -T_{{\rm KK11}} 
\int d^{7}\hat{\xi}e^{-\hat{\phi}} 
\sqrt{|{\rm det}\ (\hat{g}_{\hat{\imath}\hat{\jmath}} 
+{\hat {\cal F}}_{\hat {\imath} \hat {\jmath}})|} +{\rm WZ}\, ,
\end{equation}

\noindent which is precisely the action for the D-6-brane. Our results
suggest the identification

\begin{equation}
T_{{{\rm KK11}}} = T_{{\rm D6}}\, .
\end{equation}

%%%%%%%%%%%%%%%%%%%%%%%%%%%%%%%%%%%%%%%%%%%%%%%%%%%%%%%%%%%%%%%%%%%%

\subsection{Double Dimensional Reduction}

We next perform a double dimensional reduction of the KK11
action. We only give the reduction for the purely gravitational part.
Thus, our starting point is

\begin{equation}
S^{\rm grav.} = -{T_{KK11}\over 2}
\int d^{7}\hat{\xi}\sqrt{|\hat{\gamma}|}\ \left[
\hat{\hat{k}}^{4/7} \hat{\gamma}^{\hat{\imath}\hat{\jmath}} 
\hat{D}_{\hat{\imath}}\hat{\hat{X}}^{\hat{\hat{\mu}}}
\hat{D}_{\hat{\jmath}}\hat{\hat{X}}^{\hat{\hat{\nu}}}
\hat{\hat{g}}_{\hat{\hat{\mu}}\hat{\hat{\nu}}} -5 \right]\, ,
\end{equation}

\noindent where the covariant derivative is defined as

\begin{equation}
\hat{D}_{\hat{\imath}}\hat{\hat{X}}^{\hat{\hat{\mu}}}
=\hat{\partial}_{\hat{\imath}}\hat{\hat{X}}^{\hat{\hat{\mu}}}
+ \hat {C}_{\hat \imath} \hat{\hat{k}}^{\hat{\hat{\mu}}}\, .
\end{equation}

Usually, in a double dimensional reduction a
worldvolume coordinate $\sigma$ and a target-space coordinate $y$
associated to an isometry are simultaneously eliminated using an
ansatz of the form

\begin{equation}
Y(\hat{x}^{\hat i})=\sigma\, ,  
\end{equation}

\noindent while the other coordinates $\hat{X}^{\hat{\mu}}$
are independent of $\sigma$ so

\begin{equation}
\partial_{\sigma}  \hat{\hat{X}}^{\hat{\hat{\mu}}} =
\delta^{\hat{\hat{\mu}} y}\, .
\end{equation}

\noindent In the present case it is natural 
to make an ansatz that respects the gauge
symmetry of the gauged sigma-model. We therefore take

\begin{equation}
\hat{D}_{\sigma}  \hat{\hat{X}}^{\hat{\hat{\mu}}} =
\delta^{\hat{\hat{\mu}} y}\, .
\end{equation}

However, this would eliminate the component $\hat{C}_{\sigma}$ and for
consistency we have to make sure that this is consistent
with its algebraic equation of motion

\begin{equation}
\hat{\hat{k}}_{\hat{\hat{\mu}}} \hat{D}_{\sigma}
\hat{\hat{X}}^{\hat{\hat{\mu}}} =0\, .
\end{equation}
This implies that we must take

\begin{equation}
\hat{\hat{k}}_{y}=0\, .  
\end{equation}

\noindent For simplicity, we furthermore take

\begin{equation}
\hat{C}_{\sigma}=0\, .
\end{equation}

We now split the
$(6+1)$-dimensional worldvolume metric as follows\footnote{The
  worldvolume indices are $\hat{\imath}=(i,\sigma)$ so
  $\hat{\xi}^{6}=\sigma$}:

\begin{equation}
\left\{
\begin{array}{rcl}
\hat{\gamma}_{ij} & = & \ell^{-1/2}\gamma_{ij} -\ell^{2}a_{i}a_{j}\, , \\
& & \\
\hat{\gamma}_{i\sigma} & = & -\ell^{2}a_{i}\, , \\
& & \\
\hat{\gamma}_{\sigma\sigma} & = & -\ell^{2}\, , \\
\end{array}
\right.
\hspace{1cm}
\left\{
\begin{array}{rcl}
\hat{\gamma}^{ij} & = & \ell^{1/2}\gamma^{ij}\, , \\
& & \\
\hat{\gamma}^{i\sigma} & = & -\ell^{1/2}a^{i}\, , \\
& & \\
\hat{\gamma}^{\sigma\sigma} & = & -\ell^{-2}+l^{1/2}a^{2}\, . \\
\end{array}
\right.
\end{equation}
Substituting our ansatz for the coordinate $Y$ and for the worldvolume
metric into the action we get

\begin{equation}
\begin{array}{rcl}
S^{\rm grav.} & = & -{\textstyle\frac{T_{KK11}}{2}}
\int d^{6}\xi d\sigma \sqrt{|\gamma|}\ 
\left\{
\hat{\hat{k}}^{4/7}
\left[
\gamma^{ij}
D_{i}\hat{X}^{\hat{\mu}}D_{j}\hat{X}^{\hat{\nu}}
\hat{\hat{g}}_{\hat{\mu}\hat{\nu}} 
\right.\right.\\
& & \\
& & 
\left.\left.
-2a^{i}
D_{i}\hat{X}^{\hat{\mu}}
\hat{\hat{g}}_{\hat{\mu}y}
-(\ell^{-2}-\ell^{1/2}a^{2})
\hat{\hat{g}}_{yy}
\right] 
-5 
\right\}\, , \\
\end{array}
\end{equation}

\noindent where we have used

\begin{equation}
\hat{\hat{k}}^{\hat{\mu}} = \hat{k}^{\hat \mu}\, ,
\hspace{.5cm}
\hat{C}_{i}=C_{i}\, ,
\Rightarrow
\hat{D}_{i}\hat{X}^{\hat{\mu}}=D_{i}\hat{X}^{\hat{\mu}}\, .
\end{equation}

We next eliminate the $\sigma$-components of the worldvolume metric
($\ell,a_{i}$) by using their algebraic equations of motion and obtain

\begin{equation}
S^{\rm grav.} = -{T_{KK11}\over 2}l_{11}
\int d^{6}\xi\sqrt{|\gamma|}\ 
\left[
\hat{k}^{2/3} e^{-\frac{2}{3}\hat{\phi}}
\gamma^{ij}
D_{i}\hat{X}^{\hat{\mu}}D_{j}\hat{X}^{\hat{\nu}}
\hat{g}_{\hat{\mu}\hat{\nu}}-4 
\right]\, , \\
\label{eq:KK10Pol}
\end{equation}

\noindent where we have used the relation between the 11-dimensional 
metric and the 10-dimensional metric and dilaton

\begin{equation}
\begin{array}{rcl}
e^{\frac{4}{3}\hat{\phi}} & = &- \hat{\hat{g}}_{yy}\, ,\\
& & \\
\hat{g}_{\hat{\mu}\hat{\nu}} & = & e^{\frac{2}{3}\hat{\phi}}
\left[\hat{\hat{g}}_{\hat{\mu}\hat{\nu}} 
-\hat{\hat{g}}_{\hat{\mu}y}
\hat{\hat{g}}_{\hat{\nu}y}/\hat{\hat{g}}_{yy}\right]\, ,
\end{array}
\end{equation}

\noindent and the following relation between the 11- and 10-dimensional
Killing vectors\footnote{It is in this identity (and only in this
  identity) where the condition $\hat{\hat{k}}_{y}=0$ plays a role.}

\begin{equation}
\hat{\hat{k}}^2 = e^{-\frac{2}{3}\hat{\phi}}\hat{k}^{2}\, .  
\end{equation}

\noindent Furthermore, we define

\begin{equation}
l_{11}=\int d\sigma\, .  
\end{equation}

The action (\ref{eq:KK10Pol}) is the purely gravitational part of
the KK10A monopole action and, thus, we find

\begin{equation}
T_{KK11}l_{11}=T_{KK10A}\, .  
\end{equation}

Eliminating from the action the worldvolume metric $\gamma_{ij}$ by
using its equation of motion, we get

\begin{equation}
S = -T_{KK10A} \int d^{6}\xi e^{-2\hat{\phi}} \hat{k}^{2}
\sqrt{{\rm det}\ |D_{i}\hat{X}^{\hat{\mu}}D_{j}\hat{X}^{\hat{\nu}}
\hat{g}_{\hat{\mu}\hat{\nu}}|}\, ,
\end{equation}

\noindent which shows that both the factor $e^{-2\hat{\phi}}$
characteristic of a solitonic object and the factor $\hat{k}^{2}$
characteristic of a KK monopole are present in front of the
kinetic term. 

%%%%%%%%%%%%%%%%%%%%%%%%%%%%%%%%%%%%%%%%%%%%%%%%%%%%%%%%%%%%%%%%%%%%

\section{$T$-duality}

As a final piece of evidence in favour of the KK11 action we consider
$T$-duality. More explicitly, we will show that the heterotic
Kaluza-Klein mo\-no\-pole (KKh) is $T$-dual to the solitonic
five-brane (P5h)\footnote{The $T$-duality between KK monopole and
  five--brane solutions corresponding to the effective action has been
  considered in the context of the magnetic ciral null model
  \cite{kn:CT} and $p$-brane bound states \cite{kn:CP}.}. Before
doing this we first wish to comment on the Buscher's $T$-duality rules
\cite{kn:B}. The standard derivation of the Buscher's rules goes via a
worldsheet duality transformation on the isometry scalar in the string
effective action. A disadvantage of this derivation that it only
applies to strings but not to five-branes since the dual of a scalar
is a scalar only in two dimensions.  However, it turns out that there
is an alternative way of deriving the Buscher's rules which is more
suitable for our purposes. Combining the fact that a wave is $T$-dual
to a string and that the corresponding source terms are given by a
massless particle and a string, respectively, one can show that the
massless particle is $T$-dual to the string via reduction to d=9
dimensions \cite{prep}. This way of formulating Buscher's $T$-duality
is identical to the way the type II $T$ duality between the $R\otimes
R$ fields is treated \cite{kn:BHO}\footnote{An interesting feature of
  this alternative derivation of Buscher's rules is that the duality
  rule of the dilaton is needed already at the classical level.}.  It
is in this sense that we show below that the KKh and P5h actions are
$T$-dual to each other.

Our starting point is the 10-dimensional heterotic five-brane (P5h)
action given by

\begin{eqnarray}
S_{\rm P5h} = -T_{\rm P5h} \int d^6\xi e^{-2\hat\phi}\sqrt{ |{\rm det} \ 
(\partial_i{\hat X}^{\hat\mu}\partial_j{\hat X}^{\hat\nu}
{\hat g}_{\hat\mu\hat\nu}) |}\  + {\rm WZ}\, , 
\end{eqnarray} 

A direct dimensional reduction of the P5h action leads to
an action involving an extra worldsheet scalar $S$:

\begin{equation}
S_{\rm P5h} = - T_{\rm P5h} 
\int d^6\xi\ e^{-2{\phi}} k^{-1} \sqrt {|{\rm det}\ (  
\partial_i X^{\mu}\partial_j{X}^{\nu}
g_{\mu\nu} - k^2F_iF_j)|} + {\rm WZ}\, ,
\end{equation}

\noindent with

\begin{equation}
F_i = \partial_i S - A_i\, .
\end{equation} 

On the other hand, the heterotic KK monopole action KKh is given by

\begin{equation}
S_{{\rm KKh}} = - T_{{\rm KKh}} 
\int d^6\xi\ e^{-2{\hat\phi}} {\hat {k}}^2 \sqrt {|{\rm det} ( \ 
\partial_i{\hat X}^{\hat\mu}\partial_j{\hat X}^{\hat\nu}
{\hat \Pi}_{\hat\mu\hat\nu} - {\hat {k}}^{-2}
{\hat F}_i{\hat F_j})|} +{\rm WZ}\, , 
\end{equation}

\noindent with

\begin{equation}
\hat {F}_i=\partial_i {\hat S}
-\partial_i {\hat X}^{\hat\mu}{\hat k}^{\hat\nu}
{\hat B}_{\hat\mu\hat\nu}\, .
\end{equation}

A reduction of the KKh action over the $z$-direction gives

\begin{equation}
S_{\rm KKh} = - T_{{\rm KKh}} 
\int d^6\xi\ e^{-2{\phi}} k \sqrt {|{\rm det} ( \ 
\partial_i X^{\mu}\partial_j{X}^{\nu}
g_{\mu\nu} - k^{-2}F^{\prime}_iF_j^{\prime})|} +{\rm WZ}\, .
\end{equation}

\noindent Furthermore, we have
 
\begin{equation}
F^\prime _i = \partial_i S - B_i\, .
\end{equation}  

Combining the above reductions we see that the P5h and KKh actions
reduce to two actions in nine dimensions that differ by the following
interchanges:

\begin{equation}
 k \leftrightarrow k^{-1}\, , \hskip 1truecm
A\leftrightarrow B\, ,
\end{equation}

\noindent which are exactly Buscher's rules in nine-dimensional 
language \cite{BKO}.  This proves the $T$-duality between the P5h and
KKh actions.

%%%%%%%%%%%%%%%%%%%%%%%%%%%%%%%%%%%%%%%%%%%%%%%%%%%%%%%%%%%%%%%%%%%%

\section{Discussion}

We have proposed an effective action for the 11-dimensional 
Kaluza-Klein monopole that is given by a gauged sigma-model. 
We collected the following pieces of evidence in favour
of this proposal:

\begin{enumerate}

\item The KK11 action is the source of the KK11 solution.

\item The action reduces to the D-6-brane action via direct
  dimensional reduction and gives the KK10A action via double
  dimensional reduction.

\item The truncated 10-dimensional KKh action is T-dual to the P5h
  action.

\end{enumerate}

\noindent
It should be not too difficult to derive the Wess-Zumino term and 
to include it in the calculations performed in this letter \cite{prep}. 

It is natural to consider the
kappa-symmetric extension of the KK11 action. In fact a natural ansatz
for the relevant projection operator is (omitting the double hats)

\begin{equation}
\Gamma = {1\over 7! \sqrt {|\gamma|}} \epsilon^{i_1\cdots i_7}
D_{i_1} X^{\mu_1}\cdots D_{i_7} X^{\mu_7}\Gamma_{\mu_1\cdots \mu_7}\, .
\end{equation}

\noindent This form of the projection operator leads to the 
following Killing spinor condition for unbroken supersymmetry
\cite{BKOP}:

\begin{equation}
(1 - \Gamma_{01\cdots 6})\epsilon = 0\, .
\end{equation}
This result for the Killing spinor
can be used as another argument for a 6-brane
interpretation of the 11-dimensional Kaluza-Klein monopole
\cite{deroo}.

One might also wonder whether our results shed new light on the
evasive 11-dimensional 9-brane (see also \cite{kn:H}). 
A standard argument against the
9-brane is that the corresponding 10-dimensional worldvolume field
theory does not allow multiplets containing a single scalar to
indicate the position of the 9-brane. A way out of this is to assume
that the 9-brane is really a 8-brane with an extra isometry in one of
the 2 transverse directions, leading to a gauged sigma-model. Now, we
are dealing with a nine-dimensional field theory which naturally
contains a vector multiplet with a single scalar. In this context we
note that the 11-dimensional origin of the D-2-brane requires a
worldvolume duality transformation
of the Born-Infeld (BI) vector into a scalar
\cite{town}.  This only works for massless backgrounds, i.e.  $m=0$,
due to the presence of a topological mass term for $m \ne 0$.
However, for $m\ne 0$ it is still possible to dualize {\sl on-shell}
leading to the following line element for the eleventh scalar:

\begin{equation}
\partial_i X^{11} - m V_i\, .
\end{equation}
In other words, the general line element is given by

\begin{equation}
\partial_i X^\mu - V_i k^\mu\, ,
\end{equation}

\noindent with $k^\mu = m\delta^{\mu 11}$. We thus end up with something
which is similar to the line element of the gauged sigma-model we
considered in this letter:

\begin{equation}
\partial_i X^\mu - C_i k^\mu\, .
\end{equation}

It would be interesting to pursue this line of thought further and see 
whether it leads to a proper formulation of the long sought for 
11-dimensional 9-brane.

Finally, we believe that gauged sigma-models will play a relevant role as
source terms for many purely gravitational solutions of 11-dimensional
supergravity. The gauging procedure generically seems to allow for new
potentially supersymmetric effective actions without the need for higher
dimensions.

%%%%%%%%%%%%%%%%%%%%%%%%%%%%%%%%%%%%%%%%%%%%%%%%%%%%%%%%%%%%%%%%%%%%%%%

\section*{Acknowledgments}

T.O.~is most grateful to C.~G\'omez for useful conversations, to
M.M.~Fern\'andez for encouragement and to the Institute for
Theoretical Physics of the University of Groningen for their kind
hospitality and financial support in the early stages of this work.
B.J.~thanks the Theory Division of CERN for its hospitality. 
This work is part of the research program of the ``Stichting voor
Fundamenteel Onderzoek der Materie'' (FOM). It is also supported by
the European Commission  TMR programme ERBFMRX-CT96-0045, in which
E.B.~is associated to the University of Utrecht, and by a
NATO Collaborative Research Grant.

%%%%%%%%%%%%%%%%%%%%%%%%%%%%%%%%%%%%%%%%%%%%%%%%%%%%%%%%%%%%%%%%%%%%%%%

\end{document}